\documentclass[prd,12pt,showkeys]{revtex4-1}

\usepackage{amsfonts,amsmath} 
\usepackage{mathrsfs} 
\usepackage{setspace}
\usepackage[utf8]{inputenc} 
\usepackage{comment} 
\usepackage{hyperref}

\newcommand{\be}{\begin{equation}} 
\newcommand{\ee}{\end{equation}}

\newcommand{\goesto}{\rightarrow}

\newcommand{\lie}[1]{{\mathscr L}_{\raisebox{-1mm}{$\scriptstyle #1$}}}

\begin{document}

\title{Killing Horizons and Spinors}

\author{Bruno Carneiro da
Cunha\footnote{\footnotesize bcunha@df.ufpe.br}}

\affiliation{Departamento de Física, Universidade Federal de
Pernambuco,
50670-901, Recife, Pernambuco, Brazil}

\author{Amilcar de
Queiroz\footnote{\footnotesize amilcarq@unb.br}}
\affiliation{Instituto de Física,
Universidade de Bras\'ilia, Caixa Postal 04455, 70919-970, Brasília,
DF,Brazil}

\abstract{We study the near horizon geometry of generic Killing horizons
constructing suitable coordinates and taking the appropriate scaling
limit. We are able to show that the geometry will always show an
enhancement of symmetries, and, in the extremal case, will develop a
causally disconnected ``throat'' as expected. We analyze the implications
of this to the Kerr/CFT conjecture and the attractor mechanism. We are
also able to construct a set of special (pure) spinors associated with the
horizon structure using their interpretation as maximally isotropic
planes. The structure generalizes the usual reduced holonomy manifold in
an interesting way and may be fruitful to the search of new types of
compactification backgrounds.} }


\preprint{\today}

\maketitle

\section{Introduction}

The emergence of the supergravity equations of motion as the requisite
of conformal invariance of the worldsheet of the superstring
\cite{Polchinski:1998,Friedan:1985ge} may be one
of the most powerful ideas proposed by string theory. It addresses at
once the problem of background independence of the superstring
\cite{Friedan:1985ey}, if
partially, and provides a geometrical view of the renormalization group
(RG) flow of a number of theories which have $\sigma$-models as an
effective description at some point in coupling space
\cite{Polyakov:1987}. Also, it provides
a definition of the ``off-shell'' string as a $\sigma$-model away
from the conformal fixed point \cite{Witten:1986qs,Zwiebach:1992ie}. 

The RG flow of two-dimensional field theories is moderately well-known and their fixed
points are known to follow from maximization principles
\cite{Zamolodchikov:1986gt}. The geometrization of
the flow provided by string theory allows one to treat one of the
directions of the target manifold as a ``cut-off'' (Liouville) direction,
directly controlling the energy of the exchanges in the effective field
theory description. In this sense, one can treat the RG flow as
``displacement'' in that Liouville direction, with the effective field theory
living in the transverse dimensions.

One can then tackle, using geometrical tools, the RG flow of a variety of
quantum field theories -- at least those which arise from a background
of string theory. Versions of the ``c-theorem'' in higher dimensions
\cite{IntriligatorNucl.Phys.B667:183-2002003,Tachikawa2006} were 
the first application of this idea. Later, it was used to argue for an
appearance of a mass gap in thermal backgrounds in supersymmetric QCD
\cite{Witten1998b}. In this geometric context, the 
predictability of the RG flow stems in large part from certain algebraic
structures of the target space. These structures can be used to ``protect'' sectors
of the dual theory and talk about fixed points at the end of the flow
process. In the geometrical description (here dubbed the ``gravity side''), the RG
flow process occurs in the ``radial direction'', and has two natural
endpoints: i) the fixed point of some rotational symmetry (``$r=0$''), or
ii) an event horizon of a Black Hole. Studying the near horizon
geometry of the black hole then becomes necessary for a description of
the theory at the IR fixed point.

The foremost example of such algebraic structures arise in
compactifications that preserve $N=1$ supersymmetry in four dimensions
\cite{IntriligatorNucl.Phys.B667:183-2002003,Tachikawa2006}. 
This requisite imposes that the compactified space has a complex structure
of the Calabi-Yau type \cite{Berenstein2007}. One can then use arguments
based on supersymmetry to predict that certain properties of the theory at
the ultra-violet will not be spoiled at low energies. Two questions arise
at this point: i) whether non-renormalization theorems can be generalized
to -- possibly -- non-supersymmetric backgrounds, and ii) which
geometrical features of the target space are necessary in order to claim
that a sector will be protected under the RG flow.

From the ``gravity side'', such structures have been less well studied.
Supersymmetric structures have deep ties with special holonomy manifolds
\cite{Gualtieri2007},but the geometrical condition for no mass gap is
weaker. For instance, the first version of the Kerr/CFT correspondence
\cite{Guica:2008mu,Kunduri2007,Compere:2012jk} hinted that a local conformal symmetry
should occur for any extreme (zero-temperature) black hole. In fact, the
global part of the symmetry was found in an earlier work by Bardeen and
Horowitz \cite{Bardeen:1999px}. The Kerr/CFT correspondence showed that
the local version of the symmetry provides a suitable ``gauge slice'' of
the diffeomorphism invariance, which can be assigned to physically
independent, macroscopic distinguishable degrees of freedom. Much of the
initial properties were model-dependent, but later work by the authors
\cite{daCunha:2010jj} showed that only basic facts, like the laws of black
hole thermodynamics (BHT), were necessary to construct the conformal
symmetry, again in the extremal case.

The present work explores further this result from the $\sigma$-model side
(``gauge side''), locating the type of algebraic structures responsible
for the protection of (part of) the spectrum under the RG flow.  Some
geometrical results are needed along the way. In particular, we will show
that generically (i.e. independently of the Lagrangean of the theory),
the set of conserved charges associated with remaining symmetries --
including Kaluza-Klein charges -- are set on the horizon independently of
their value at infinity. We will dwell on these issues in sections 2, 3
and 4. Furthermore, their boundary values satisfy a maximum principle,
stemming from the first law of BHT. This can been interpreted as a
geometrical verification of the ``attractor mechanism''
\cite{Sen2007,AstefaneseiJHEP0610:0582006,0264-9381-23-21-S04,Hotta2009,
Larsen2008}.  

The latter part of the paper (section 5 onwards) is dedicated to the
structures responsible for the existence of the protected sector. In order
to relate them with the known reduced holonomy case, we will need to
present some elementary notions of algebraic geometry and spinors (section
5). We tie the discussion (section 6) with some prospects for the
construction of relevant backgrounds.

\section{Killing Horizons}

As stated above, the endpoint of the RG flow process in the geometrical
picture may be, in general, an event horizon of a black hole. In this section
we will give a geometrical description of this fact, and derive an analogue of
the zeroth law of the black hole thermodynamics (BHT) in a
Lagrangean-independent way.  

The event horizon of a black hole is formally defined as the boundary
of the causal past of an observer sitting at (null or spacelike) infinity \cite{Wald}.
Intuitively, this notion entails the idea that, far enough in the
future, the geometry is sufficiently ``settled'' (strong asymptotic
predictability) that one can speak about a particular region of space. 

As always, in order to be easily workable, the notion of event horizon
presupposes the existence of a symmetry of the spacetime. Thus, as it is
usual, we consider for the sake of workability event horizons that are
Killing horizons as well. Observe however that this strategy is not of
itself without hurdles. In fact, even the existence of conserved charges
like the mass and angular momentum is not automatic in general relativity.
It also asks for the existence of suitable symmetries like time
translation and rotation around an axis in order to be unambiguously
defined. 

Much of the ensuing discussion will follow an essay by Carter
\cite{Carter1973}, with a few extra added results which are relevant for
our construction. Newer treatments can be found in
\cite{Ashtekar:2001jb} and \cite{Pawlowski:2003ys}. To the best of our
knowledge, a derivation of the laws of black hole thermodynamics in
generic dimensions based solely on the concept of Killing horizon and
integrability like the one we present here is new. Of course,
knowledge of some specific form of the dynamics will be necessary to
study the near-horizon geometry, and the conclusion that the notion of
algebraic speciality plays an important role there was anticipated by
the work of R. Milson and collaborators, as for instance
\cite{Milson:2004jx}. 

Let us begin by recalling the usual four-dimensional case. 

We denote by $k^a$ and $m^a$ the time-translation and axial rotation Killing vector
fields (KVFs), respectively. By default, these vectors are
tangent to integral lines of functions $t$ and $\phi$ and are
commuting $[k^a,m^b]=0$. The event 
horizon will be defined as the region where a combination of these vectors
becomes null. From the commuting condition, we have that
\be
k_a=-V(dt)_a+W(d\phi)_a,\quad\quad m_a=W(dt)_a+X(d\phi)_a
\ee
The event horizon is then the region where a combination of the inner products
become null, that is, $V=k^ak_a$, $X=m^am_a$ and $W=k^am_a$ are such that 
\be
\rho = VX+W^2\rightarrow 0,
\ee
where $\rho$ can be thought of as the determinant of the restriction
of the metric to the plane formed from $k^a$ and $m^a$. When $\rho$
becomes zero, this plane becomes null 
and finite acceleration timelike curves cease to exist in it. The most
important result \cite{Wald} that follows is that the combination of $k^a$
and $m^a$
that becomes null at this point,
\be
\chi^a=k^a+\Omega_H m^a,
\ee 
is actually also a Killing vector field. To see that, define
$\Omega_H=-W/X$ and consider
\be
X^2d\Omega_H=-XdW+WdX.
\ee
Note that 
\be
X^2d\Omega_H\wedge dt\wedge d\phi = X^2d\Omega_H\wedge k
\wedge m = \rho^2dm\wedge m.
\ee
Now, given that the far right hand side vanishes, then $d\Omega_H\wedge k
\wedge m=0$. This means that $d\Omega_H$ is parallel to $k^a$ and $m^a$, but
this is not possible since $k^a$ and $m^a$ are Killing vector fields and
$\Omega_H$ is a scalar function constructed out of the metric. The
quantity $\Omega_H=-W/X$ is thus constant throughout the horizon. It
corresponds to its angular velocity. 

The vector $\chi^a$ is orthogonal to the horizon: the axial
rotation vector $m^a$ is normal to it. Also, since it is orthogonal to
a surface of constant $\rho$, it is also hypersurface orthogonal
there, satisfying Frobenius' condition
\be
\chi\wedge d\chi = \chi_{[a}\nabla_b\chi_{c]}=0 \quad \text{ at the
horizon.}
\ee
We will see that this condition poses a powerful constraint on the
geometry near the horizon in the general case.

The conditions above are suitably generalized to higher dimensional
spacetimes. 

Consider an even $D$-dimensional spacetime with $D/2$ commuting
(dual) KVFs, denoted by $\{\Psi_i\}$. This setup
corresponds to the physical situation where one has a ``vacuum'' spacetime
which allows for $D/2$ independent conserved quantities. In more
dimensions they will add up to the total mass, the (Weyl subalgebra of)
the angular momenta, as well as charges arising from Kaluza-Klein
compactifications, be they abelian or non-abelian.

In general, one considers the problem of finding $D/2$ coordinates 
$\{\phi^i\}$ covering the orbits of each of those KVFs. In these
coordinates, the vectors associated to the $\Psi_i$ will be
$\partial/\partial\phi^i$. The condition
that such $\{\phi^i\}$ exist is actually a zero curvature problem. Indeed, if
there are $\lambda_{ij}$ functions such that
$$
\Psi_i=\lambda_{ij}d\phi^j,
$$
then 
\be
d\Psi_i={\mu_i}^k\wedge \Psi_k.
\label{frobenius}
\ee
The ``connection'' ${\mu_j}^k$ is actually a flat one, since the above
two equations imply that
$$
{{\cal F}_j}^k=d{\mu_j}^k+{\mu_j}^l\wedge {\mu_l}^k=0,\quad\quad
{\mu_j}^k=(\lambda^{-1})^{kl}d\lambda_{lj}.
$$
The vanishing of ${{\cal F}_j}^k$ can be interpreted as an existence
condition for the $\lambda_{ij}$ (also known as flat
parametrization).The equation \eqref{frobenius} is called 
Frobenius' theorem for dual (covariant) vectors. In contrast to the usual
connections, the $\lambda_{ij}$ are the components of the matrix of inner
products between the $\Psi_i$ and are actually symmetric, and thus they do
not form a group. Specifically, the separation between the
orbits of the KVFs $\{\phi^i\}$ and the transverse space form a sort
of a bundle which is not a principal one. 

\subsection{Compatible Killing Vector Fields}

Defining the killing horizon as the region where one combination of the
KVFs become null allow us to probe the near horizon geometry. In essence,
the situation is not much different from the Euclidian case where any
isometry near a fixed point can be seen as a rotation. 

A null eigenvector of $\lambda_{ij}$ implies that one combination
of the 1-forms $\Psi_i$ becomes null.  As such the coordinates
$\{\phi^i\}$ are not suitable for describing the horizon itself. Since
$\lambda_{ij}$ is symmetric, one can decompose it as
\be
\lambda= z P + X,
\label{lexpansion}
\ee
where $z$ is a selected eigenvalue, which the one going to zero at the
horizon, $P$ is the (orthogonal) projector to the would-be ($z=0$) null 
eigenspace, and $X=(1-P)\lambda(1-P)$ is the restriction of $\lambda$ to the
orthogonal subspace.

Define a particular combination of KVFs 
\begin{equation}
\chi=\Omega^i\Psi_i=\Omega_i(d\phi^i), \qquad \textrm{with } 
\Omega_i=\lambda_{ij}\Omega^j.
\end{equation}  
If $\chi$ is to be a KVF, then
the $\Omega^i$ have to be constant. Now, consider
\be
\begin{aligned}
\chi\wedge d\chi & = -\frac{1}{2}(\Omega_i d\Omega_j -\Omega_j
d\Omega_i)\wedge d\phi^i\wedge d\phi^j \\
& = -\frac{1}{2}(\Omega_i d\Omega_j-\Omega_j
d\Omega_i)(\lambda^{-1})^{ik} (\lambda^{-1})^{jl}\wedge \Psi_k \wedge
\Psi_l,
\label{hypers}
\end{aligned}
\ee
the last step being necessary since the coordinates $\{\phi^i\}$ are not
all valid at the horizon. The inverse of \eqref{lexpansion} is
$$
\lambda^{-1}=\frac{1}{z}P+Y,\qquad\text{where}\quad Q\equiv XY=YX=1-P.
$$
Substituting this back to the last line of \eqref{hypers} and using 
that both $\chi$ and $\Psi_i$ are well-defined at $z\goesto 0$, we
find that
\be
P^{ik}(\Omega_i d\Omega_j-\Omega_j d\Omega_i)P^{jl}={\cal
O}(z^2),\quad\quad P^{ik}(\Omega_i d\Omega_j-\Omega_j
d\Omega_i)Y^{jl}={\cal O}(z).
\label{hyperorth}
\ee
The first condition is trivial if the rank of $P$ is
one. Let us consider this case first, and comment
about what happens in general. Assuming that $Y^{ij}$ is non-degenerate on
the orthogonal subspace, the second condition implies
\be
P^{ik}(\Omega_i d\Omega_j-\Omega_j d\Omega_i)Q^{jl}={\cal
O}(z),
\ee
which, specializing to the case where $\chi$ is actually the null KVF at
the boundary -- we can take $P$ to project onto the first component, and
$Q$ onto the others -- results in
\be
d\left(\frac{\Omega_i}{\Omega_1}\right)={\cal O}(z)
\label{0thlaw}
\ee
which means that the $\omega_i=\Omega_i/\Omega_1$ are {\em constant over
the horizon}. Apart from the temperature, which we discuss later, this
implies the zeroth law of BHT, 
and does not depend on details of the dynamics, just on the presence
of a Killing horizon. 

If the rank of $P$ is greater than one, there will be families of
conserved intrinsic quantities,
$$
\Omega_\mu d\Omega_i-\Omega_i d\Omega_\mu =\Omega_\mu^2
d\left(\frac{\Omega_i}{\Omega_\mu}\right)={\cal O}(z)
$$
where $\mu$ ranges over the image of $P$ and $i$ over the orthogonal
complement. This alters little the discussion that follows, but the
differences, especially when defining coordinates, will be outlined in
the appropriate sections.

Going back to the case where $P$ is of rank one, the conditions
\eqref{hyperorth} can be solved by requiring that all $d\Omega_i$ are
parallel, that is, there is a function $\beta$ such that $d\Omega_i=\Omega_i d\beta
+{\cal O}(z)$. Thus,
\be
\nabla_a\chi_b = d\Omega_i\wedge d\phi^i = d\beta \wedge \Omega_i
(d\phi^i)
+{\cal O}(z) = d\beta \wedge \chi + {\cal O}(z),
\label{hsochi}
\ee
that is to say, $\chi$ is {\em hypersurface orthogonal} at the horizon ($z=0$).
By contracting the equation above with $\chi$ itself, one arrives at
\be
\beta = \log |\chi^a\chi_a| + {\cal O}(z).
\ee

So far we have only shown that the ``potentials'' $\omega_i$ are
constant over the horizon. We will postpone the discussion about the
``temperature'' $\kappa$ to the next subsection. With this provision, the
discussion above shows that the existence of a Killing
horizon gives enough constraints on the ``intrinsic'' quantities (like
the temperature, angular velocities and others associated with
Kaluza-Klein charges) that render them constant over the horizon. We have
thus arrived at a Lagrangean independent formulation of the zeroth law of
BHT, as anticipated by Carter.

\subsection{Hypersurface Orthogonal Killing Vector Fields}

Let us discuss \eqref{hsochi} in more detail. Suppose that we
have a spacetime with a Killing vector field (KVF)
$\chi^a$. 
Let us assume, for simplicity, that this KVF is hypersurface orthogonal, 
$\chi_{[a}\nabla_b\chi_{c]} =0$. The preceeding discussion shows that
this holds ``to order $z$'' near the Killing horizon, so every
equation in this section will be valid up to ${\cal O}(z)$. 

From the Killing equation $\nabla_{(a}\chi_{b)}=0$, we obtain the identity
\begin{equation}
	\label{Killing-identiy-1}
  \nabla_a\nabla_b\chi_c=R_{cbad}\chi^d.
\end{equation}
Now, Frobenius' theorem implies that 
\begin{equation}
  \chi^a R_{a[b}\chi_{c]}=0.
\end{equation}
Einstein manifolds ($R_{ab}=\frac{R}{D}g_{ab}$) satisfy naturally this
condition, but of course this condition is less restrictive.  

Given the KVF $\chi^a$, we define its norm $z$ and its gradient
$N_a$, respectively, as  
\be
\chi^a\chi_a=z,\quad\quad
N_a=\frac{1}{2}\nabla_a z=-\chi^c\nabla_c\chi_a. 
\ee 
We then note that $N^a$ is basically the acceleration of the integral
curves of $\chi^a$ with $N^a\chi_a=0$. The norm of $N^a$ is
\be 
N_a N^a=-\kappa^2 z.
\ee 
Later we interpret $\kappa^2$ as the ``surface gravity''. We note that
$\kappa$ is imaginary if both $\chi^a$ and $N^a$ are space-like. 

As we assumed that the KVF is hypersurface orthogonal, its covariant
derivative can be calculated readily \cite{Wald} as  
\be
\nabla_a\chi_b=-\frac{\chi_{[a}\nabla_{b]}(\chi^c\chi_c)}{\chi_d\chi^d}
=-\frac{2}{\alpha}\chi_{[a}N_{b]}. \label{derivchi} 
\ee 
This, in turn, means that 
\be
(\nabla^a\chi^b)(\nabla_a\chi_b)=\frac{2}{z^2}(\chi^a\chi_a)(N_bN^b)
=-2\kappa^2 \label{secondkappa} 
\ee 
The normalized bivector normal to surface levels of $z$,
\begin{align} 
\epsilon_{ab} =\frac{1}{\kappa}\nabla_a\chi_b &=-\frac{2}{\kappa
z}\chi_{[a}N_{b]}, \qquad \textrm{with} \\
 \epsilon^{ab}\epsilon_{ab} &=-2, 
\end{align}
is a purely geometrical quantity which is well-defined at the horizon
${\mathscr H}$ ($z=0$) even in the extremal limit $\kappa\to 0$. Using the
identity (\ref{Killing-identiy-1}), we can also compute the covariant
derivative of $N_a$,  
\be 
\nabla_a N_b=-\frac{\kappa^2}{\alpha}
\left(\chi_a\chi_b-\frac{1}{\kappa^2}N_aN_b\right)+
\chi^cR_{cabd}\chi^d~=~-\kappa^2\epsilon_{ac}{\epsilon^c}_b+ 
\chi^cR_{cabd}\chi^d. \label{lieg}  
\ee 
From the above facts, we can compute with a little effort, the
following formulas,
\begin{align} 
\label{kappa-square-1}
\nabla_{a}\kappa^2 &=-\kappa\chi^bR_{abcd}\epsilon^{cd}=~\Sigma N_a-\kappa
\omega_{ab}\chi^b, \\
\nabla_{[a}\epsilon_{bc]} &=-\frac{1}{2\kappa^2}\epsilon_{[ab}\nabla_{c]}
\kappa^2+\frac{1}{\kappa}R_{[cba]d}\chi^d=~-\frac{1}{2\kappa}\epsilon_{[ab}
\omega_{c]d}\chi^d, 
\label{somederiv} 
\end{align} 
where in the final step, Bianchi identity $R_{[abc]d}=0$ was used, and the
quantities $\Sigma$ and $\omega_{ab}$ are defined in terms of the decomposition 
\be
R_{abcd}\epsilon^{cd}=-\Sigma\epsilon_{ab}+\omega_{ab}; \quad\quad
\omega_{ab}=\omega_{[ab]},\quad \text{and}\quad
\omega_{ab}\epsilon^{ab}=0. 
\label{definitions}
\ee 
We thus see that $\Sigma$ is associated with the area of the plane defined
by $\epsilon_{ab}$ (it may be thought of as the generalization of the
expansion parameter of congruences of curves) $\omega_{ab}$ is associated
with the parallel transport perpendicular to this plane (it may be thought
of as the generalization of the rotation or twist parameter of congruences
of curves).  

Movement normal to the surfaces with constant $z$ is generated by a
vector parallel to $N^a$, properly normalized, so that 
\be
n^a\nabla_a z=1\longrightarrow n^a=-\frac{1}{2\kappa^2 z}N^a.
\label{affinen} 
\ee 
We think of $n^a$ as $\partial/\partial z$. Then, with the above
formulas, we can compute the Lie derivative of $\epsilon_{ab}$ with
respect to $n^a$ -- recall that $\lie{\xi}{\epsilon} = \xi\cdot d\epsilon
+ d(\xi\cdot\epsilon)$--, arriving at  
\be
\lie{n}{\epsilon_{ab}}=-\frac{\Sigma}{4\kappa^2}\epsilon_{ab}+\frac{1}{
4\kappa^2}
\left(\omega_{ab}+\frac{2}{z}\chi_{[a}\omega_{b]c}\chi^c\right)
=-\frac{\Sigma}{4\kappa^2}\epsilon_{ab}\label{lieepsilon}. 
\ee 
Note that the vanishing of the term between brackets is a result of the Bianchi
identity $R_{[abc]d}=0$. From the derivative of $\kappa^2$,
(\ref{kappa-square-1}), we obtain  
\be 
\lie{n}\kappa^2=\frac{1}{2}\Sigma. 
\label{liekappa}
\ee 
This, along with (\ref{lieepsilon}), 
means that the tensor $\kappa \epsilon_{ab}$ is independent of $z$.
We will use this fact in the next subsection to derive the approximate
form of the metric near the horizon. For now, we consider the term inside
the brackets in \eqref{lieepsilon}, rewritten as
\be
\chi_{[a}\omega_{b]c}\chi^c=-\frac{z}{2}\omega_{ab}
\ee
which can be translated to 
\be
\omega_{ab}\chi^b\propto \chi_a+{\cal O}(z).
\ee
This has a deep implication to \eqref{kappa-square-1}. The
derivative of the surface gravity $\nabla_a\kappa$ points in the
$\epsilon_{ab}$ plane, up to terms of order $z$. It is thus {\em
constant over the horizon}. Later we will see that $N_a$ actually becomes
proportional to $\chi_a$ at the horizon, but for now this fact suffices
to finish the proof on the zeroth law of BHT.

Let us now introduce the null vectors
\be
\xi^a_\pm=\frac{1}{\kappa}N^a\pm\chi^a.
\label{nullvects}
\ee
After some algebra, we may show that
\be
\nabla_a(\xi_\pm)_b=-\kappa \epsilon_{ac}{\epsilon^c}_b+\frac{1}{\kappa}
\left(\delta^e_b+\frac{1}{\kappa^2 z}N^eN_b\right)\chi^cR_{caed}
\chi^d\pm \kappa\epsilon_{ab}
\label{xiderivative}
\ee
and then
\begin{align}
\label{xiderivatives-1}
\xi^b_\pm\nabla_b\xi^a_\pm &=-2\kappa\xi^a_\pm -z h^{ab}\nabla_b\kappa,\\
\xi^b_\mp\nabla_b\xi^a_\pm &=-z h^{ab}\nabla_b\kappa,
\label{xiderivatives-2}
\end{align}
where $h_{ab}=g_{ab}-{\epsilon_a}^c\epsilon_{bc}$ is the ``angular
part'' of the metric. The last term in the above equation projects out
the $N^a$ component of  the derivative of $\kappa$. 

Let us define the vectors
\be
l^a=\frac{1}{2}\xi_+^a \qquad \textrm{ and }\qquad n^a={\epsilon^a}_b\xi_+^b,
\ee
so that both have well-defined limits as we take $z\rightarrow
0$. Such limit is easily computed for $l^a$. Indeed, since $N^a$ becomes
null at the horizon, and it is always orthogonal to $\chi^a$, it must
be proportional to $\chi^a$ itself. From the definition of $\kappa$,
we can easily see that
\be
l^a=\chi^a\quad\quad\text{at the horizon}.
\ee

The limit of $n^a$ is a little more subtle. From its definition (\ref{xiderivative}),
it can be thought of to be proportional to $\xi^a_-$, but in such a way
that it does not vanish at the horizon. Indeed, recall that 
$\epsilon_{ab}=2n_{[a}l_{b]}$ everywhere. From the definition of
$\xi^a_\pm$, (\ref{nullvects}), we obtain
\be
n^a=\frac{1}{\alpha}\left(\frac{1}{\kappa}N^a-\chi^a\right)
\quad\quad\text{at the horizon}.
\ee 
In other words, it is a vector normal to the horizon in such a way
that $n^al_a=-1$. 

Since $\kappa$ is constant at the horizon ${\mathscr
H}$, we find, by inspecting \eqref{xiderivative}  that $l^a$ is tangent to null
geodesics if $\kappa\neq 0$ at the horizon. The ``angular'' component
of the derivative of $\kappa$ will tell us the transverse acceleration
of these null geodesics. 

Looking at (\ref{kappa-square-1}), we learn that the condition
that these null vector fields are geodesic is equivalent to the
vanishing of $\omega_{ab}$. When this tensor vanishes, the canonical
bivector $\epsilon^{ab}\propto \xi_+^{[a}\xi_-^{b]}$ satisfies 
\be
R_{abcd}\epsilon^{ab}=-\Sigma \epsilon_{cd}.
\label{eigenbivector}
\ee
In other words, $\epsilon_{ab}$ is an eigen-bivector of the Riemann
tensor. It can be easily checked that for Einstein spaces,
$R_{ab}=\frac{R}{D}g_{ab}$, $\epsilon_{ab}$ is also an eigen-bivector of
the Weyl tensor. In fact, 
using the decomposition of the Riemann tensor, we have for Einstein spaces
that
\be
\Sigma=\frac{1}{2}\epsilon^{ab}\epsilon^{cd}R_{abcd}=-\frac{2}{n(n-1)}R+
\frac{1}{2}\epsilon^{ab}\epsilon^{cd}C_{abcd}.
\ee 
Further, by inspecting the form of the derivative of $l^a$ and $n^a$,
\eqref{xiderivative}, we can see that
\be
R_{abcd}l^bl^d=\frac{\Sigma}{4} l_a l_c,\quad\quad R_{abcd}n^b
n^d=\frac{\Sigma}{4} n_a n_c,
\quad\text{ at the horizon.}
\label{weylaligned}
\ee
Therefore both null directions are {\em repeated principal null
directions}. Principal null directions arise naturally in the
algebraic classification of the Weyl tensor, which in turn was
important for the hunt of solutions of General Relativity in four
dimensions. Incidentally, the conditions in \eqref{weylaligned} in four
dimensions are known as the Bel criteria \cite{Stephani:2003b} for type
$D$ solutions, and happen for all known Kerr-Newman solutions. In
higher dimensions, the notion of algebraic speciality is a bit more
involved \cite{Coley:2004jv}, and vectors satisfying
\eqref{weylaligned} are sometimes called Weyl-aligned null directions
\cite{Milson:2004jx}. The detailed structure of Weyl-aligned null
directions in higher dimensional Killing horizons can be found in
\cite{Pravda:2007ty} and \cite{Ortaggio:2009sb}.

\subsection{Null Vector on the Near Horizon: Extremal Case}

We now turn to the problem of defining suitable null vectors near the
horizon of an extremal black hole. 

Consider the derivative of
$\xi^a_{\pm}$ \eqref{xiderivative}, from which we obtain 
\be
l^a\nabla_al^b=-\kappa l^b-\frac{1}{2}zh^{bc}\nabla_c\kappa.
\label{ellderivative}
\ee 
We first consider second term on the right hand side which is proportional to the Riemann tensor
$\nabla_a\kappa = \frac{1}{2}\chi^bR_{abcd}\epsilon^{cd}$. Near
the horizon \eqref{eigenbivector} we have 
\be
\chi^aR_{abcd}\chi^d=-\frac{\Sigma}{2\kappa^2}N_bN_c+\ldots,
\ee
where we left out terms at higher order in $z$. At the horizon,
$N^a=\kappa\chi^a$ are parallel to each other, so that
\be
\chi^aR_{abcd}\chi^d=-\frac{\Sigma}{4}\chi_b\chi_c+{\cal O}(z).
\ee
Therefore $\chi^a$ is a doubly-repeated principal null vector. From this
we conclude that the second term in the right hand side of
\eqref{ellderivative}, that is, $h^{bc}\nabla_c\kappa$ is of order
${\cal O}(z^{1/2})$, and then 
\eqref{ellderivative} 
\be
l^a\nabla_al^b=-\kappa l^b+{\cal O}(z^{3/2}). \label{limitnablaell-0}
\ee
In the extremal ($\kappa\goesto 0$) limit, only terms of highest order are kept. This
means that $l^b$, after the limit is taken, can be seen as a
geodesic vector field. Moreover, the curvature-dependent term of
\eqref{xiderivative} can also be seen to vanish in the limit, being
itself of order $z^{3/2}$. Then, after the limit is taken,
\be
\nabla_a l_b = -\Gamma \rho n_a l_b.
\label{limitnablaell}
\ee
This means that the null vector $l_b$ is hypersurface orthogonal, that is,
$l_{[a}\nabla_b l_{c]}=0$ even away from the horizon. The null vector is
a principal null direction, which vanishing expansion, shear and twist. 

The usefulness of defining such a null vector in the near horizon of
an extremal black hole stems from the following property. Let $m^a_i$
and $\bar{m}^a_i$ be null (complex) vectors spanning the space
orthogonal to $l^a$ and $n^a$, constructed in such a way that the
commutators between $m^a_i$ and $l^a$ vanish. Parametrize a generic
vector in the orthogonal space by 
\be
\eta^a=\bar{z}^im^a_i+z^i\bar{m}^a_i.
\ee 
Upon parallel transport under
$l^a$, we obtain
\be
l^a\nabla_a z^i=-\rho_{ij} z^j+\sigma_{ij} \bar{z}^j,
\ee
where $\rho_{ij}=m^a_i\bar{m}^b_j\nabla_al_b$ and
$\sigma_{ij}=m_i^am_j^b\nabla_al_b$ both vanish due the expression of the derivative of
$l^a$ in \eqref{limitnablaell}. This means that one can define a
complex structure in the subspace generated by the $m_i^a$ and
$\bar{m}_i^a$ and \eqref{limitnablaell} will guarantee that such
structure can be parallel-transported throughout the variable. 

The above parallel-transported structure is related, but weaker, than
that of SUSY, which asks for the existence of a covariantly constant
spinor. We will digress in the following section about the
similarities and differences of both SUSY and the above.  

In the following sections we will discuss more about the geometrical
interpretation of these conditions. We just note that in four dimensions, the
null vectors $l^a$ and $n^a$ satisfying \eqref{weylaligned} are related
to integrability properties.

The laws of BHT state that $\kappa$ is constant over the horizon
$\mathscr{H}$, which in turn means that the tensor $\omega_{ab}$ vanishes,
and $\epsilon^{ab}$ is a principal bivector. However, we will see that the
two null vectors $\xi^a_\pm$ degenerate on ${\mathscr H}$ to $\pm \chi^a$
and thus fail to define distinct principal null vectors. If
$\omega_{ab}=0$ hold throughout, then the structure outlined in this
section will carry on to the whole manifold. 

\section{The Near Horizon Geometry}

In this section we will state several conditions on the relations
obtained above in the case where the KVF becomes null, $z\rightarrow
0$. This condition holds for stationary black holes, and in several
cases may be used as the telltale signal of an event horizon.  

\subsection{Coordinates}

At first one can think of the construction above as happening
throughout the spacetime. We will show, however, that a significant
simplification happens on the structure near a Killing horizon
${\mathscr H}$, where
$z=0$. 

We have assumed that the KVF $\chi^a$ is hypersurface
orthogonal, its covariant derivative being given by (\ref{derivchi}). 
From this, it is readily verified that the 1-form $\frac{1}{z}\chi_a$ is
closed. Then, locally, there is a function $u$ such that
\be
\chi_a=-z(du)_a.
\ee
We can think of $u$ and $z$ as local coordinates on the
spacetime, parameterizing the ``non-angular'' directions, that is,
they span the analogue of the ``$r-t$'' plane. In these coordinates,
the bivector $\epsilon_{ab}$ has the form 
\be
\epsilon_{ab}=\frac{1}{\kappa}(dz)_{[a}(du)_{b]}.
\label{canonicalepsilon}
\ee
This form is suitable to study the behavior near the horizon
$z=0$. Now, (\ref{liekappa}) and (\ref{lieepsilon}) allow us to
expand both $\kappa$ and $\epsilon_{ab}$ near $z=0$. Thus,
\be
\label{kappa-square-2}
\kappa^2(z)=\kappa_0^2+\frac{\Sigma_0}{2}z+{\cal O}(z^2), \\
\ee
where the subscript $0$ means evaluation at the horizon. These
expressions can be used to expand the metric $g_{ab}$ near the region
$z=0$. We split the metric as the semi-direct sum of the ``$r-t$''
plane and the angular coordinates, that is,
\be
g_{ab}=\epsilon_{ac}{\epsilon^c}_b+h_{ab}.
\ee
The ``$r-t$'' plane has line element
\be
ds^2=dr^2-f(r)^2du^2,
\ee
with $r(z)$ a function chosen so that $dr$ is the normalized
vector parallel to $N_a$. Also, we have $z=f(r)^2$, and then,
\be
N_a=-\frac{1}{2}dz = -f(r)f'(r)dr,\quad
N^aN_a=-f^2(r)[f'(r)]^2=-\kappa^2z.
\ee 
The function $f(r)$ is then determined from the
knowledge of $\kappa$ as
\be
\quad \kappa = |f'(r)|, 
\ee
which is known near $z=0$. We can then distinguish two cases.

\subsubsection{Case $\kappa_0>0$}

This is the best known case. Since the laws of BHT tell that
$\kappa_0$ is constant at the horizon, we 
have $\lim_{r=0}|f(r)|=\kappa_0$ and then $f(r)=\kappa_0r$ for small
$r$. The line element is then 
\be
ds^2=-dr^2+\kappa_0^2r^2du^2,
\ee
which shows the well known property that an analytic continuation for
imaginary $u$ will display a conical singularity unless $u$ is
identified with period $2\pi/\kappa_0$. We will borrow some
terminology -- however appropriate -- and dub this case ``elliptic''.

The 2-form $\epsilon_{ab}$ has the canonical form from
(\ref{canonicalepsilon}),
\be
\epsilon_{ab}=\frac{1}{\kappa_0}(dz)_{[a}(du)_{b]},
\ee
which can be easily used to define a canonical volume form on a
topological sphere that intersects ${\mathscr H}$, as in the first law
of BHT.

\subsubsection{Case $\kappa_0=0$}

This is the case of foremost interest to us. It is dubbed
``hyperbolic''. If $\kappa_0=0$, then we have to go to next order in
$z$ in (\ref{kappa-square-2}),
\be
\kappa^2=\frac{\Sigma_0}{2}z=\frac{\Sigma_0}{2}f(r)^2.
\ee
This gives us the following differential equation for $f(r)$:
\be
f'(r)=\Gamma f(r)\qquad \Rightarrow \qquad f(r)=\exp(\Gamma r),
\ee
with $\Gamma^2 = \Sigma_0/2$. The metric is then
\be
ds^2=-dr^2+\exp(2\Gamma r)du^2,
\ee
which has constant negative (sectional) curvature. 

The volume form $\epsilon_{ab}$ can be cast as
\be
\epsilon_{ab}=\frac{1}{\Gamma}\frac{(dz)_{[a}(du)_{b]}}{z^{1/2}},
\ee
so that the canonical variable is now $\rho=z^{1/2}$.

Note that in both cases the isometry of the ``$r-t$'' plane has been
enlarged (or enhanced). 
Instead of just a line of symmetries coming from the KVF, they are now
symmetries of the Minkowski plane ${\rm SO(1,1)}$ in the case of
positive $\kappa_0$ and the ubiquitous ${\rm SL(2,\mathbb{R})}$ in the
hyperbolic case. Whether the action of those new generators keep the
full metric is not known.

Along with the above two cases, there is another where
$\kappa_0=\Sigma_0=0$, 
which we call ``parabolic''. It has also a canonical form of the
metric, but no enlarged symmetry or constant curvature.

\section{The Bottomless pit}

In this section we study the near-horizon geometry in the extremal
case. We show that the enhancement of symmetries discussed
in the previous section can be carried over to the near horizon
region. Also, we show that the near horizon region becomes causally
disconnected with the asymptotic region.

Very close to the horizon $z=0$, the integral curves of $\xi^a_\pm$ come
closer and closer to being geodesics. From the equation
\eqref{xiderivative}, one sees that the affine parameters $w^\pm$ of such
geodesics are related to the affine parameters $x^\pm$ of
$\xi^a_\pm$ by 
\be
w^\pm = \exp\left[\int^{x^\pm}dx'\,\kappa\right].
\ee
From this we can again distinguish two cases. If
$\kappa\goesto\kappa_0\neq 0$ at the horizon, then $w^\pm$ has a
minimum value, and then the horizon is incomplete in the
past. Geodesics which asymptote the horizon as we take $x^\pm\goesto 
-\infty$ will be incomplete in the past as well, because for geodesics
sufficiently close to the horizon the affine parameter can be
approximated by $w\approx e^{\kappa_0x}$. At the horizon, the point on
which $w^\pm=0$ is actually a fixed point of the Killing vector field
$l^a$, and it is called {\it bifurcation point} \cite{Townsend1997a}. 

The situation changes somewhat when $\kappa\goesto 0$ at the
horizon. Now we can take $w^\pm=x^\pm$ and the geodesics can be
indefinitely extended. Introducing coordinates such that
$\chi^a=\partial/\partial u$ and $\rho^a=-2\kappa^2
z \partial/\partial z$, then $\xi^a_\pm$ is given 
by
\be
\xi^a_\pm=-2\kappa z \frac{\partial}{\partial z}\pm\frac{\partial}{\partial u}.
\ee
Given that $\kappa\goesto 0$ at the horizon, the affine parameter of
$\xi^a_\pm$ will depend crucially on the behavior of $\kappa$ close to the
horizon. By \eqref{liekappa}, we have the expansion
\be
\kappa(z)^2=\kappa_0^2+\frac{1}{2}\Sigma_0z+{\cal O}(z^2)
\ee
where the subscript $0$ refers to quantities being computed at the
horizon. Integrating \eqref{kappa-square-2} with $\kappa_0=0$ for the
affine parameters $x^\pm$, we have 
\be
x^\pm=\pm u + A z^{1/2}=\pm u + A \rho.
\ee
These have two interesting features. First and foremost, unlike the
case where $\kappa_0\neq 0$, the curves can be continued indefinitely,
meaning that the region near the black hole has a causal infinity --
the ``throat is bottomless'' or ``the bottomless pit'' \cite{Carter1973}
-- which can be thought of as a  
decoupling limit of the induced theory near the boundary and the
asymptotic region far from the horizon. We will have more to say about
the holographic interpretation below.

The second feature can be seen as a consequence of the first. One
notes that, in order to focus in the near-horizon $z\approx 0$ region,
one can make a scale transformation, which for $\kappa_0\neq 0$,
involves the logarithm of the coordinate $z$,
\be
\xi^a_\pm \goesto \lambda^{-1} \xi^a_\pm \Longrightarrow u\goesto \lambda
u\quad\text{ and }\quad z\goesto z+A^{-1}\log \lambda.
\ee
For $\kappa_0=0$, the transformation is different,
\be
\xi^a_\pm \goesto \lambda^{-1} \xi^a_\pm \Longrightarrow u\goesto \lambda
u\quad\text{ and }\quad \rho\goesto \lambda\rho.
\ee
This last transformation allows for a holographic interpretation. The
coordinate $\rho$ actually changes scales without any dimensionful
parameter (like $\kappa_0$). We will argue below that the absence of
dimensions in the $\rho$ coordinate can be thought of as the signal of
a fixed point on the renormalization group flow.

We can now make a simple argument supporting the so-called ``attractor
mechanism'' conjectured to hold for extremal black holes
\cite{Sen2007,AstefaneseiJHEP0610:0582006}. Indeed, from
\eqref{0thlaw} we can argue the following regarding  
the values of the angular velocities $\Omega^i$ at the horizon. It was
shown that those are constant on the horizon. Now, by the argument
given above, these values at the horizon are causally disconnected from
the asymptotic region in the extremal case. Given this, one can set
values for the angular velocities -- the intrinsic quantities in BHT
-- independently from the values of these 
quantities at infinity. As argued above, the same reasoning can be
applied to the values of any field that can be obtained from Kaluza-Klein
reduction of gravity. 

\section{Such spinors at the Horizon}

In this section we discuss the previous results in the ``gauge
picture'', using integrability of null planes and their relation to
spinors to rewrite the geometrical results obtained for the near
horizon geometry in a setting suitable to discuss implications in the
field theory side of the gauge/gravity duality. The construction we
have outlined in the preceding sections made use of the notion of
Killing horizon to single out preferred null directions on the 
manifold. We will now explore the relationship of generic null
(isotropic) subspaces and spinors. The goal is to derive a
condition on these subspaces compatible with extremality. 

Consider a Lorentzian, even $D=2n$-dimensional manifold with real Vielbeine
$e^i$. From those we can construct a null basis $\{l^i,n^i\}$ satisfying
\be
l^i\cdot l^j=n^i\cdot n^j=0,\quad\quad l^i\cdot n^j=\delta^{ij}.
\ee
Now, some of the elements of this basis will be complex, so we will
consider the complexified tangent space arising from generic complex
combinations of $\{l^i,n^i\}$. This construction can be thought of as
splitting the tangent space at each point into an isotropic vector space
$V$ and its dual $V^*$: $T_pM\simeq V\oplus V^*$. Observe that
$\{l^i\}$ forms a basis of $V$ and $\{n^i\}$ forms a basis of $V^*$
The ``natural'' pairing between $V^*$ and $V$ is given by the metric,
with the direct sum structure stemming from the linearity of the inner
product and the isotropy of $V$ and $V^*$, that is,
\be
\langle (v_1+w_1),(v_2+w_2) \rangle = v_1\cdot w_2+w_1\cdot
v_2, \quad \textrm{ with } v_i \in V, ~ w_i\in V^*.
\ee
Anyway, the particular choice of null
basis is not relevant for the following discussion, since any two of them
are related by an element of ${\rm SO}(2n)$. As algebras, the splitting is
\be
{\rm so}(V\oplus V^*)={\rm End}(V)\oplus \wedge^2 V \oplus \wedge^2 V^*,
\label{rotdecomp}
\ee
involving particular two forms (exterior products) of the $\{l^i\}$ and
the $\{ n^i\}$.

Multivectors of $V^*$ are generic linear combinations of exterior products
of the $\{n^i\}$,
\be
\varphi = a+a_i n^i+a_{ij}n^i\wedge n^j+\ldots.
\label{phispinor}
\ee
The elements of the full space $V\oplus V^*$ acts on those
multivectors via the geometric product, 
\be
(v+w)\varphi = v\cdot \varphi + w\wedge \varphi,
\label{geomprod}
\ee
where the contraction is always done with the first index. Repeating the
action, one gets
\be
(v+w)[(v+w)\varphi]=(v\cdot w)\varphi = \langle (v+w),(v+w) \rangle
\varphi.
\ee
The above geometric action is then implemented as the Clifford product on the
space of multivectors. In other words, isotropic multivectors can be seen
as spinors. Simple counting shows that the space of (simple) isotropic
multivectors is identified with Dirac spinors, whereas restriction to
multivectors of either odd or even degree results in chiral (Weyl)
spinors. 

Given a (non-vanishing) spinor $\varphi$, one defines its {\it
  annihilating space} $L_\varphi$ as the elements of 
$V\oplus V^*$, such that
\be
(v+w)\varphi=0.
\ee
As a direct consequence, one notes that any element of $V$ is a null
(isotropic) vector, since for any (non-vanishing) $\varphi\in V^*$,
\be
(v+w)(v+w)\varphi =  0 \qquad \Longrightarrow \qquad \langle
(v+w),(v+w) \rangle=0. 
\ee
Then, in the complexified vector space, $L_\varphi$ can have at most
(complex) dimension $n$. If this is 
the case, $L_\varphi$ is called {\it maximally isotropic} and $\varphi$
is called a {\it pure spinor}\footnote{A note on jargon
  \cite{Dubois-Violette2007}. The term pure spinors was set by
  C. Chevalley, while E. Cartan dubbed it simple spinors. These pure
  spinors span a Fock space \cite{Dubois-Violette2007}. This last fact
  lies at the heart of the importance of these objects.}
\cite{Cartan1981}. One such spinor is $\varphi = 1$, which is
annihilated by all elements of $V$ (via \eqref{geomprod}). Therefore
$L_1$ has maximal dimension. 

A known result \cite{Cartan1981} states that all pure spinors are related by
the action of an element of ${\rm SO}(2n)$. So, to any given maximally
isotropic space $L$, one can find a single (up to scaling) pure spinor
$\varphi_L$ such that $L$ is its annihilating space. The idea is that $L$
can be ``rotated'' to $V$ by a suitable element of ${\rm SO}(2n)$, and
then this element will bring $\varphi_L$ to $1$. Given the decomposition
\eqref{rotdecomp}, this element is unique up to endomorphisms of $L$,
which leave $\varphi_L$ invariant up to scaling.

We may now consider the introduction of a connection on the manifold,
so that we can parallel transport the above structures over the
manifold. For that we should deal with the geometric product using
Leibniz rule  
\be
\nabla(v\varphi)=(\nabla v)\varphi + v\nabla\varphi,
\ee
which has to vanish if $v$ is an element of $L_\varphi$. Requiring that
the right hand side vanishes imposes that $\nabla v$ has only components
along $L_\varphi$ and $\nabla\varphi$ is proportional to $\varphi$.

The condition that the splitting $T_pM\rightarrow V\oplus V^*$ holds
throughout the manifold is now easily seen as either of the equivalent
notions:
\begin{enumerate}
 \item Parallel transport of any vector in $V$ stays into $V$, that is,
the holonomy is reduced from ${\rm SO}(2n)$ to ${\rm SU}(n)$.
 \item There is a parallel, covariantly constant, (pure) spinor,
associated via \eqref{phispinor} to a maximally isotropic plane.
\end{enumerate}
Either condition is related to remaining supersymmetry. The construction
makes sense in the complexified tangent space, so the choice of ``real
form'' (whether the signature is Euclidian or Lorentzian) has to be
compatible with an integrability structure. In practice this means that
the SUSY charges satisfy the correct reality condition leading to the correct
algebra, so that they anticommute to the Hamiltonian. 

As we have stressed throughout, all the discussion applies when the
integrable null plane is maximal. 

In the preceding sections we built the argument that one has (necessarily)
a principal null direction on a Killing horizon. This will be necessarily
real, and can be extended to the near-horizon in the extremal case. Extra
commuting Killing vector fields related to ``transverse'' symmetries will
also be able to generate similar structures. The word ``transverse''
merely states that the conserved quantities arising from the symmetries
are constructed uniquely from the symplectic structure (cf. discussion in
\cite{Iyer:1994ys}). In the maximal case, we can define $D/2$ null, orthogonal
but not necessarily real lines bundled to a maximally
isotropic plane. From the preceding discussion, these are associated with a parallel
spinor. One can then see that the construction resembles very closely the
relation between reduced holonomy and SUSY, although the remark about
signature can prevent the charges thus defined to play a role in
organizing the spectrum of the theory. 

In four dimensions, the conditions arising from the existence of
integrable maximal 
isotropic planes result in familiar structures. The Goldberg-Sachs
theorem \cite{Goldberg:2009rp} 
relates the existence of integrable isotropic planes to repeated principal
spinors. One assigns null vectors belonging to a null plane by 
\be
\ell^a=\sigma^a_{\alpha\dot{\alpha}}\iota^{\alpha}\bar{\pi}^{\dot{\alpha}},
\ee
with $\sigma^a_{\alpha\dot{\alpha}}$ the van der Waerden symbols (also
chiral Pauli matrices or soldering form). To relate with the discussion above, we may
think of a spinor basis $\{\iota^\alpha,o^\alpha\}$ generating the
two-dimensional $V$. We will omit the identification between spinors and
vectors and write $\ell^{\alpha\dot{\alpha}}$ for a vector from now on.
The spinor $\bar{\pi}^{\dot{\alpha}}$ is generic, but the plane contains
the real vector $\iota^\alpha\bar{\iota}^{\dot{\alpha}}$. The condition
of integrability is 
\be
\iota^\beta\iota^\alpha\nabla_{\alpha\dot{\alpha}}\iota_\beta=0.
\ee
This merely states that any commutator of vectors formed from 
$\iota^{\alpha}\bar{\pi}^{\dot{\alpha}}$ belongs to the same plane. The
condition can be trimmed slightly since one can freely rescale the
spinor, so that 
\be
\iota^\alpha\nabla_{\alpha\dot{\alpha}}\iota^\beta=0.
\label{pspinor}
\ee
Now, using the definition of the curvature spinor,
\be
\nabla_{\dot{\alpha}(\alpha}{\nabla^{\dot{\alpha}}}_{\beta)}
\kappa_\gamma =
\Psi_{\alpha\beta\gamma\delta}\kappa^\delta-2\Lambda\kappa_{(\alpha}
\epsilon_{\beta)\gamma},
\ee 
one arrives from \eqref{pspinor} at an algebraic condition involving the
spinorial version of the Weyl tensor $\Psi_{\alpha\beta\gamma\delta}$, that is, 
\be
\Psi_{\alpha\beta\gamma\delta}\iota^\alpha\iota^\beta\iota^\gamma=0.
\label{pspinor2}
\ee
The spinor space is two-dimensional, hence
$\pi_{\alpha}\iota^{\alpha}=0$ implies that $\pi^\alpha$ and
$\iota^\alpha$ are proportional. Furthermore,
$\Psi_{\alpha\beta\gamma\delta}$ can be factorized into the principal
spinors:
$\Psi_{\alpha\beta\gamma\delta}=\kappa^1_{(\alpha}\kappa^2_{\beta}
\kappa^3_{\gamma}\kappa^4_{\delta)}$. The condition \eqref{pspinor2}
means that $\iota^\alpha$ will appear twice in this decomposition.
So the spinor associated with the null plane is necessarily repeated. We
can then work out two cases, whether there are one or two integrable null
planes, and the Weyl tensor is either one of the following two forms, 
\be
\Psi_{\alpha\beta\gamma\delta}=A\iota_{(\alpha}\iota_{\beta}
\iota_{\gamma}\iota_{\delta)}, \quad\quad\text{ or }\quad\quad 
\Psi_{\alpha\beta\gamma\delta}=A\iota_{(\alpha}\iota_{\beta}o_{\gamma}o_{
\delta)}.
\ee
These correspond to the Petrov type N and D Weyl tensor, respectively.
Associated with the spinor(s) one can assign 
\be
\Phi_{\alpha\beta}=A\iota_{\alpha}\iota_{\beta}\quad\quad\text{ or }\quad\quad
\Phi_{\alpha\beta}=A\iota_{\alpha}\iota_{\beta}+A' o_{\alpha}o_{\beta}.
\ee
One can then construct the Killing-Yano antisymmetric tensor
$G_{ab}=G_{[ab]}=\Phi_{\alpha\beta}
\epsilon_{\dot{\alpha}\dot{\beta}}+\text{c.c.}$. At this point the
coefficients $A$ and $A'$ are generic functions. The Killing-Yano tensor
satisfies $\nabla_{(a}G_{b)c}=0$ and its existence is also tied to
integrability properties of the space-time. In fact, the integrability
structure is exactly the same as outlined above. Killing-Yano tensors have
been important in the integration of the geodesic equation of type D
solutions \cite{Walker:1970rp,Krtous:2006qy} and in the separability
of the wave equations in those backgrounds
\cite{Frolov:2006pe,Mason:2010ky}.  

In more than four dimensions, however, one can encounter a lower
dimensional integrable null plane which cannot be seen as the real or
imaginary part of a maximal one. A spinor associated to such plane is
called {\it impure}. In fact, a lower dimensional null plane is not
associated to a single spinor, but with a number of them. One can write
schematically
\be
\varphi = \sum_i a_i\psi_i,
\ee
where $\psi_i$ are pure spinors, but in general $\varphi$ is not. The
coefficients $a_i$ are arbitrary. Geometrically, the $\varphi$ is
associated to a null plane of annihilators $L_\varphi$ which is the
intersection of the annihilators of each of the pure spinors $L_{\psi_i}$.
One can now state the condition that $L_\varphi$ is integrable by
demanding that the set of pure spinors $\{\psi_i\}$ satisfy the involutive
property 
\be
\nabla\!\!\!\!\raisebox{2pt}{/}\,\psi_i=\sum_j a_{ij}\psi_j.
\label{involutive}
\ee 
This condition is milder than SUSY at six or more dimensions. 

The conclusion of the above discussion is that {\em Killing horizons have associated
with them a set of pure spinors $\{\psi_i\}$ satisfying the involutive
property \eqref{involutive}. When the horizon has zero temperature, the
property can be extended to the near-horizon geometry.} The number of pure
spinors in the set reflects the dimension of the integrable null plane: in
the one-dimensional case one has $d/2-1$ pure spinors whose intersection
gives a particular null line. When the set of pure spinors consists of
just one element, the geometry of the spacetime is of reduced holonomy and
one can use the standard arguments to show that SUSY charges can be
defined. In the general case, however, the condition seems milder and may
be of relevance in the quest for backgrounds for AdS/CFT, among other
applications. 

We will close the section by discussing examples in 6 dimensions. The
notation chosen when writing \eqref{involutive} tries to make clear that
$\{\psi_i\}$ should be thought of as Dirac spinors. Let us work explicitly
with the basis for spinors in 3 complex dimensions, 
\be
\{1,n_1,n_2,n_3,n_1\wedge n_2,n_1\wedge n_3,n_2\wedge n_3,n_1\wedge n_2\wedge n_3\}.
\ee
Each element of the basis is a pure spinor. For example, $L_1=\{
\ell_1,\ell_2,\ell_3\}$. Any sum of same chirality -- Weyl -- spinors
either has dimension $3$ or $1$, like $n_1+n_1\wedge n_2\wedge n_3$, whose
annihilator is $n_1$. In order to obtain a two-dimensional plane, one has
to consider the sum of terms with opposite chirality: the annihilator of
$1+n_1$ is, for instance, $\{\ell_2,\ell_3\}$. The condition
\eqref{involutive} may then involve the sum of spinors of different
chiralities. It would be interesting to understand its relation to usual
constructions of supersymmetry and supergravity.

\section{Discussion}

Let us now apply the geometrical construction outlined to the case
where the dual field theory has a sigma-model description. From the
geometrical side, the horizon is, by definition, a surface of last contact. It is
usually the limit of coordinates considered to be ``natural'' by an
asymptotic observer. This surface is necessarily null, and by the usual
results repeated in the preceeding sections it is also integrable. 

Given an integrable null surface in spacetime, we can either associate it
with a (perhaps partial) splitting of spacetime into isotropic spaces
$V\otimes V^*$, or associate it with a number of null spinors via the
involutive property \eqref{involutive}. The first approach was crucial to
the development of special solutions in general relativity, including the
Kerr-Newman family, and ultimately to the twistor programme. The second
has deep ties to supersymmetry, which played a role akin to integrability
-- perhaps ``amenability'' would be a more suitable word -- in field
theory. We will argue that they should be thought of as equivalent
physicaly as well as mathematically.

Let us begin with the case where there is a maximally isotropic null
integrable plane. By definition, this means that the splitting
$T_pM\rightarrow V\oplus V^*$ can be done consistently throughout the
manifold. For Euclidian signature, such spaces with special holonomy are
called Kähler manifolds. As a rule, the existence of this special holonomy
combined with some restriction on the geometry -- like, for instance,
the manifold being Einstein ($R_{ab}=\frac{R}{D}g_{ab}$) is enough to
preclude non-trivial quantum corrections to spring up in the
higher-order effective theory. In fact, if the scalar curvature is
zero (and hence the second Chern class also vanishes), the manifold
has a well-defined spinor charge and is in fact a supersymmetric
background. The geometric spinors defined above can be integrated to
generate a spinorial charge. The usual non-renormalizability theorems
\cite{SeibergPhys.Lett.B318:469-4751993,SeibergNucl.Phys.B435:129-1461995,Seiberg2007a}
can be invoked to protect the potential from generating a mass gap, at
least perturbatively.  

To sum up, the constancy of spinors can then be related to the
constancy, or integrability, of the splitting between a maximally
isotropic space $V$ and its dual. Heuristically, the integrability of
the splitting is in turn related to the absence of quantum corrections
to the geometry. 

This property can be really appreciated from the
gauge/gravity perspective. Suppose we have a two-dimensional
sigma-model with target manifold $M$. Usually the geometry of $M$ is
set in the ultraviolet scale of the sigma-model and the geometry
changes as the scale goes down. Upon certain conditions, the scale
itself can be seen as an extra dimension
\cite{FreedmanAdv.Theor.Math.Phys.3:363-4171999,Bianchi2001,Bianchi2002}. In
some special cases, the renormalization group (RG) flow equations can
be seen as the truncation of the Einstein equations for the extended space. 

The nature of such truncation seems a bit mysterious at first. The RG
flow (Callan-Symanzik) equations are of first order in the scale
parameter, whereas the Einstein Equations are of second order if we
pick as the scale parameter the coordinate spanning the extra dimension. In
fact, in the usual example of this correspondence, the extra dimension
is the radial coordinate $r$ of ${\rm AdS_5}$, so that the metric $h_{ab}$
induced in the space transverse to $\nabla_a r$ satisfies, rather
trivially,
\be
\frac{\partial h_{ab}}{\partial r}=\frac{1}{R}h_{ab}.
\ee
The extrinsic curvature $K_{ab}$ is then proportional to the metric. From
the ``other half'' of Einstein's equations, one learns that
\be
\frac{\partial K_{ab}}{\partial r}=R[h]_{ab}.
\ee
Therefore only for Einstein spaces the constraint that the extrinsic
curvature is proportional to the metric will hold throughout the
evolution in the parameter $r$. 

Our understanding is that the flow along the $r$ direction should be
understood as geodesic flow. 
It is generated by the $r^a=(\partial/\partial r)^a$ vector field, conjugate
to the gradient of $r$, $r^a\nabla_a r=1$. We will choose coordinates so
that the normalized geodesic vector field $n^a$ is parallel to $r^a$,
$r^a=Nn^a$, $n^an_a=1$. The treatment parallels that of congruences of
timelike geodesics. Note that the geodesics we are talking about are
spacelike. We define the induced metric on the leaves of constant $r$, 
\be
h_{ab}=g_{ab}-n_an_b.
\ee
The derivative of $n^a$, related to the extrinsic curvature $K_{ab}$, is defined by
\be
B_{ab}=\nabla_{b}n_{a},\qquad \textrm{so that}\qquad
K_{ab}=\frac{1}{2}\lie{n}h_{ab}=B_{(ab)}.
\ee
The interpretation of the operator $B_{ab}$ is similar to its timelike
counterpart. If one defines a basis of vectors tangent to the surface of
constant $r$ by $m_i^a$, obeying the compatibility condition
\be
\lie{n}m_i^a=n^b\nabla_bm_i^a-m_i^b\nabla_bn^a=0,
\ee
then the parallel transport by $n^a$ will be
\be
\delta m_i^a=\epsilon n^b\nabla_bm_i^a=\epsilon
m_i^b\nabla_bn^a=\epsilon {B^a}_bm_i^b.
\ee
So, as the flow through $r$ progresses, the basis $m_i^a$ will be twisted
and turned by the exponential of the operator ${B^a}_b$. If we take the
usual action form for the $\sigma$-model,
\be
S_\sigma = \int d^2x \sum_i \sqrt{g}g^{\alpha\beta}h^{ab}\partial_\alpha
m_i^a\partial_\beta m_i^a+\ldots,
\ee
then the (worldsheet) derivatives of the (spacetime) vectors $m_i^a$ play
the role of the primaries ${\cal O}_i$. The action above is of course
schematic: we do not consider fermions, and the $r$ direction arises
only at the effective action level. In the holographic RG spirit
\cite{FreedmanAdv.Theor.Math.Phys.3:363-4171999}, 
movement in the $r$ direction corresponds to changing of the energy scale of
the $\sigma$-model.

A natural candidate for the RG-flow parameter in the generic case is
the scale factor $\theta$, defined as the trace of $B_{ab}$, as found
in the decomposition 
\be
B_{ab}=\theta h_{ab}+\omega_{[ab]}+\sigma_{(ab)},
\label{decompB}
\ee
where the ``shear'' $\sigma_{(ab)}$ is the traceless part of the
extrinsic curvature. Since the role of $\theta$ is to dilate the vectors,
$m_i^b$, the above assignment seems plausible. If $\omega_{[ab]}$
or $\sigma_{(ab)}$ are not equal to zero, then the transverse metric
$h_{ab}$, seen as the couplings of the sigma model is changing not only
the scale, but also the relative couplings between the sigma-model
observables (in this case, the primaries $\partial X^a$ and
$\bar\partial X^a$). The classical geometric equations dictating the
change of $B_{ab}$ are known as the Raychaudhuri equations
\cite{Wald}. Computing the second derivative of $B_{ab}$, we have
\be
n^c\nabla_cB_{ab}=-{B_{ac}}{B^c}_b+n^cR_{acbd}n^d,\quad \text{ or } \quad
\lie{n}B_{ab}={B^c}_aB_{cb}+n^cR_{acbd}n^d,
\label{raychaudhuri}
\ee
where one can use the decomposition of the Riemann tensor into the pure trace,
traceless transverse 2-tensor $S_{ab}=R_{ab}-\frac{1}{n-1}Rg_{ab}$ and the
Weyl tensor $C_{abcd}$,
\be
R_{abcd}=\frac{2}{(n-1)(n-2)}Rg_{a[c}g_{d]b}+\frac{2}{n-2}(g_{a[c}S_{d]b}
-g_{b[c}S_{d]a})+C_{abcd}.
\ee
One can then split \eqref{raychaudhuri} into equations for each of the terms
in \eqref{decompB}:
\begin{align}
\lie{n}\theta &=
-\theta^2-\frac{1}{n-1}\left(\sigma_{ab}\sigma^{ab}-\omega_{ab}\omega^{ab}
\right)-\frac{1}{n-2}R_ { ab } n^an^b; \\
\lie{n}\omega_{ab} &= 0;
\\ 
\lie{n}\sigma_{ab}
&=\theta\sigma_{ab}+\sigma_{ac}{\sigma^c}_b-\omega_{ac}{\omega^c}_b-
\frac{1}{n-1}h_{ab}(\sigma_{cd}\sigma^{cd}-\omega_{cd}\omega^{cd} )
\quad\quad\quad\quad \nonumber \\
& \quad +h_a^ch_b^dS_{bd}+
C_{cbad}n^cn^d.
\end{align}
If $B_{ab}$ turns into a pure scale
transformation, we can say that we arrived at a fixed point of the RG
flow process, in which the primaries no longer change. One sees from the
equations above that near the horizon of a black hole which is an
Einstein manifold, the conditions for the endpoint are met because
$n^a$ effectively becomes a repeated principal null vector. Therefore
the source terms on the right hand side vanish. One should 
also note that {\it a priori} the choice of initial condition
$\omega_{ab}=0$ is maintained as one follows through the flow. In
physical terms, the tensor $\sigma_{ab}$ encodes the non-trivial
renormalization process. One could in principle have a non-trivial
$\omega_{ab}$ by introducing a ``lapse vector'' $N^a$, and then the
Lie derivatives of a generic tensor $T_{ab}$ would be modified into
\be
\lie{n}T_{ab}=\frac{1}{N}\left(\frac{\partial}{\partial r}T_{ab}-
N^c D_c T_{ab}-T_{cb}D_aN^c-T_{ac}D_bN^c\right),
\ee
where $D_a$ is the covariant derivative associated with $h_{ab}$, the
projection of $\nabla_a$ to the surfaces of constant $r$.

Coming back to the Callan-Symanzik equation, the fact that the Einstein
equations are second order entails the fact that the observables entering
the (gauge independent) correlation functions are themselves changing with
the energy scale, that is,
\be
\delta {\cal O}_i = \sum_{j} B_{ij}{\cal O}_j.
\ee
Such mixing can only happen for theories for which there are a large
number of operators sharing the same scaling dimension. One notes that
this variation mixes different correlation functions in 
the Callan-Symanzik equation and can be thought of as a higher order
correction. Of course, near a conformal
fixed point, the only change 
up to first order in the beta function is that of the scaling
dimension, corresponding to a diagonal $B_{ab}$. 
 
We have two examples illustrating the above discussion. For asymptotically
anti-de Sitter (aAdS) space-times, the metric can be written as
\be
ds^2=\frac{dr^2}{r^2}+h_{\mu\nu}(z,x^\rho)dx^\mu dx^\nu
\ee
with $h_{\mu\nu}$ approaching a conformally flat metric as we take
$r\goesto 0$. In this regime, the condition $K_{ab}=\theta h_{ab}$ is
satisfied and the usual lore says we are arriving at a (UV) conformal fixed
point. The addition of marginal or relevant perturbations -- like the
addition of a subleading mass term -- will mix the primaries as the
flow goes down. Perhaps the most famous example of this mixing happens
when we have several ${\rm U(1)}$ global symmetries. The RG flow mixes
the generators of each of the ${\rm U(1)}$ factors. If spontaneous
symmetry breaking occurs along the way, then the generators of the
remnant ${\rm U(1)}$ symmetries at the infrared fixed point may be
radically different from those proposed at the ultraviolet. In some
important applications, such generators can be found from a
variational principle -- ``anomaly maximization'', presumably for the
same reasons the algebraic structure made explicit above is relevant for
extremal black holes.

The second example comes in the guise of the many versions of the
``c-theorem'', a generalization of the famous work in two
dimensions \cite{Zamolodchikov:1986gt}. General relativity coupled to
matter satisfying the strong energy condition will always have $\theta
< 0$ as $r$  increases, moving $h_{ab}$ away from the UV point. This
is in tune with the irreversibility of the RG-flow, in which the
number of degrees of freedom decreases under the
RG-flow. Geometrically, $\theta$  counts the change of small elements
of area with $r$. The assignment of some degrees of freedom with the
area is then natural from BHT. 

\section{Conclusions \& Perspectives}

In this article we showed that the ``geometric definition'' of a
black-hole in terms of local existence of a null Killing vector field
entails a great deal of information about the local geometry. We were
able to show that an enhancement of the geometry should happen, but
whether this enhancement can be extended to the near horizon limit seems
to be feasible only in the extremal case, where it is causally
disconnected from the asymptotic region. Given the absence of a mass
scale as well as the disconnection with asymptotic observers, it is less
of a surprise that one can elect a class of gauge-inequivalent metrics
that counts the Hawking-Bekenstein entropy, in the spirit of
Kerr-CFT. 

Also, one of the major consequences of the analysis is that the
disconnection of the region near the horizon and the asymptotic is
independent of the dynamics and the dimension of the theory, and is a
direct consequence of extremality. In particular, one can then dictate
values for the global charges in the near horizon regime are
independent from those in the asymptotic region. These global charges
can be thought of basically any charge that can be obtained from a
suitable Kaluza-Klein compactification of pure gravity, so it
encompasses not only the charges associated with the Killing vector
fields, but also scalar charges, flavor charges, abelian and non-abelian
charges as well as monopoles. These are fixed from the Second Law,
which now can be stated purely from the geometry, not depending on the
details of the dynamics or supersymmetry. Hence, the construction
outlined here provides a geometrical verification of the attractor
mechanism which relies solely on the integrable structure of 
the near-horizon region.

The second part of this work dealt with the algebraic structures behind
the extremality. It is known that the near horizon limit displays an
enhancement of (super)symmetries for supergravity backgrounds. We claimed
that even in the non-supersymmetric case one deals with the involutive
property of null planes, which are naturally associated with
spinors. Thus the requirement of supersymmetry is not mandatory in the
generic case, and might as well be replaced with the integrability of
(lower-dimensional) null planes. In terms of spinors, the latter
translates into the involutive property described in
\eqref{involutive}. The same classical integrable structure which is
phenomenologically interesting for SUGRA backgrounds also allows for the
solutions of Einstein equations in pure gravity. Quantum mechanically,
these spaces should have a geometric description of their gravitational
degrees of freedom, just like in Kerr/CFT. If in four dimensions one
encounters essentially the known cases (either a Calabi-Yau or complex
charges), the situation changes in 6 dimensions, where one can have
interesting cases of non-maximal null planes
\cite{Batista:2012cp}. Also, it would be interesting to rewrite the
SUSY algebra in terms of the splitting proposed here, which we believe
should help in the search of backgrounds of 10d SUGRA. 
 
\section*{Acknowledgements}
We would like to thank Francisco Brito, Carlos Batista, A. P.
Balachandran, Javier Martinez Magan, Dmitry Melnikov for discussions and
incentive in several stages of this work. We also thank Álvaro Ferraz, the
director of IIP-UFRN in Natal, Brazil, together with the staff of this
institute for the kind assistance in the many occasions we had used the
facilities of the institute to discuss this project. BCdC thanks the
ICCMP-UnB, where part of this work was conducted. ARQ is supported by CNPq
under grant no. 305338/2012-9 and CAPES under grant no. 8713-13-8.


\end{document}